# Two Types of Natural Kind Discovery: Nobel meets Kuhn

Samuel Schindler
Centre for Science Studies
Aarhus University

**Abstract**

Philosophers have spilled much ink over the discovery of ideas in the classical "context of discovery". However, there has been little engagement with the question of what constitutes a discovery of "things in the world". A much-overlooked answer to this question is provided by T.S. Kuhn. In this paper, I show that discoveries awarded with a Nobel Prize in Physics in the past 53 years accord with a basic premise of Kuhn's account and his distinction between two types of natural kind discoveries. I also draw normative conclusions for credit attribution in science.

## 1  Introduction

Traditionally, philosophers have thought of discovery as discovery of *ideas*. A central question in classical debates about discovery has therefore been whether there is a logic to how scientists discover their ideas, theories, hypotheses, etc. (see Schickore 2022 for an overview). In contrast, the discovery of *things in the world*, and more specifically the discovery of *natural kinds* and their *properties* has received much less attention (Kuhn 1962; Achinstein 2001; Arabatzis 2006; McArthur 2011; Schindler 2015; Copeland 2019; Arfini et al. 2020; Duerr and Holmes Mills 2025).

This widespread neglect of discovery of "things in the world" seems somewhat scandalous: what makes the news and what we all get excited about are not so much the *ideas* scientists produce, but rather whether the ideas lead to something more tangible, as when the electron was discovered in 1897, gravitational light bending in 1918, the structure of DNA in 1953, or the Higgs boson in 2012. The reason for philosophers' neglect of discovery cannot be due to triviality: historians have long known that discovery is not, as Kuhn put it, like "a single simple act assimilable to our usual (and also questionable) concept of seeing" (Kuhn 1962/1996, 55). However, historians tend to overemphasize the complexity and the contingency of discovery (Brannigan 1980; Schaffer 1986; Arabatzis 2006). As Hanson put it provocatively, the work by historians "never reflect monochromatically: only spectra of concepts and arguments result" (Hanson 1962, 582).

Luckily, Kuhn did provide some structure (no pun intended!) to our conception of natural kind discovery. In chapter IV of *The Structure of Scientific Revolutions* (SSR), Kuhn argued that discovery – and his examples suggest that he meant discovery of natural kinds – involves not only observing for the first time *that* something is the case, but also *what* it is that has been observed (Kuhn 1962/1996, 55). In other words, a discovery also requires a *conceptualization* of the newly observed or detected phenomenon. Based on this basic premise, Kuhn introduced a distinction between two types of discovery: a type of discovery in which a novel phenomenon is *first* conceived of theoretically and *then* observed, and a type in which a novel, unexpected phenomenon is *first* observed and *then* conceptualized. Since the latter type of discovery cannot be accommodated in the reigning paradigm, Kuhn associated this type with scientific revolutions; the former type of expected discovery, he associated with normal science.

Kuhn's distinction between the two types of discovery has been largely ignored, possibly because the two types were simply subsumed under the two rubrics of normal science and revolutions, and possibly also because in SSR the distinction is not as clear as it is in a paper by Kuhn that appeared in the same year as SSR in *Science* magazine and which was dedicated specifically to discovery (Kuhn 1962). Interestingly, in this paper Kuhn does not at all mention the central notions of SSR of paradigm change, revolutions, or incommensurability. This goes to show that Kuhn's distinction can be treated independently from the rest of the controversial Kuhnian conceptual apparatus (Schindler 2015). Whereas Kuhn did not give any names to the two types of discovery, I have suggested earlier that we refer to them as *that-what* discoveries and *what-that* discoveries (ibid..). *That-what* discoveries are discoveries in which a surprising phenomenon is first observed and then conceptualized, whereas *what-that* discoveries are discoveries in which the discovered phenomenon is predicted on the basis of present theoretical resources. I will use this terminology in what follows.

Despite the apparent neglect of Kuhn's distinction[1], there has been some discussion of the premise which the distinction is based on, namely that discovery of natural kinds requires some substantial *and correct* conceptualization in the first place. Especially Hudson (2001) has challenged this premise (see also Achinstein (2001)). I will here not engage with these challenges to the Kuhnian premise; I have done so elsewhere (Schindler 2015). My focus in this paper is rather on empirically testing this premise and the Kuhnian distinction between two types of natural kind discovery. In order to do so, I analyzed the Nobel Prizes in Physics from 1972 to 2024. My findings support both Kuhn's basic premise and the distinction.

Why is Kuhn's distinction worth pondering for philosophers? Well, first and foremost, the question of "what is a scientific discovery of natural kinds?", at core, is an question of concept explication on a par with any other prominent questions that philosophers spend their time discussing, such as "what is the scientific method?", "what is a scientific explanation?", "what is causation?", etc. The fact that very few philosophers have hitherto tackled the question, or take the question to be trivial (mistakenly, I would argue), does not mean that the question is not a worthy philosophical one. On the contrary, given the centrality of natural kind discovery for science, I believe it would continue to be a scandal if philosophers kept ignoring the question.

As mentioned, my focus here will be on empirically testing an account of scientific discovery of natural kinds. Why should concept clarification have any business in seeking empirical adequacy? According to the standard Carnapian account, concept explication involves several desiderata, one of which is *similarity* between explicandum and explicatum, which is meant to ensure that the analysis explicate the intended concept and not something else, or even nothing at all (Carnap 1950). This is one way of viewing this project; another would be to treat philosophical theories as akin to scientific ones in that the former are just as much in need for empirical justification as the latter (Giere 1985; Donovan et al. 1988).

---

[1] Apart from my own work (Schindler 2015), Duerr and Holmes Mills (2025) have recently picked up on the distinction and developed a more expansive account of discovery that goes beyond the discovery of natural kinds (the subject of the current paper).

What's in it for those more interested in the practical application of ideas rather than in conceptual clarification per se? Concept clarification is obviously meant to alleviate conceptual confusion, which can have very concrete practical implications. In the case of discovery, confusion about what it is that constitutes a discovery of natural kinds may lead to skewed and unfair credit attribution practices. Those, in turn, may undermine the reward system of science, which has been argued to be critical for the progress of science (Strevens 2003; Zollman 2018).

The paper is structured as follows. Section 2 introduces Kuhn's distinction and presents the hypotheses for the empirical test. Section 3 expounds the method that was used in the empirical analysis of the Nobel Prizes. Section 4 reports and discusses the results obtained. Section 5 of this paper summarizes the results and draws some normative conclusions regarding current credit attribution practices.

## 2    Kuhn's Distinction and Hypotheses

Kuhn's basic premise that underlies his distinction is best motivated by the example that Kuhn himself used in SSR, namely the discovery of oxygen. Joseph Priestley was arguably the first person to isolate and notice oxygen as a separate species of gas in 1774 (and to publish it). Until the end of his life, though, he believed that he had discovered de-phlogisticated air. But that's clearly not what Priestley discovered: phlogiston does not exist. What he did discover, instead, was a gas that the phlogiston theory could not accommodate easily, namely oxygen. Thus, the discovery of oxygen did not only involve *that* there was a new gas, but also a realization about *what* that gas actually was. Moreover, since in the discovery of oxygen the new entity was *first* detected and *then* correctly conceptualized, we can speak of a *that-what* discovery.

The other type of natural kind discovery, namely *what-that* discovery, plays less of a role in both SSR and Kuhn's *Science* article; it is still more visible in the latter than in the former. Again, in this type of discovery, the newly detected phenomenon is not at all surprising because it has been predicted by theory / anticipated by the ruling paradigm. Examples that Kuhn gives of *what-that* discoveries are new chemical elements predicted by Mendeleev, and the discoveries of neutrino and radio waves (Kuhn 1962, 761; 1962/1996, 58). As Kuhn memorably put it, this kind of discovery is "occasion only for congratulation, not surprise" (Kuhn 1962/1996, 52). For example, the discovery of the Higgs particle in 2012 at the LHC at CERN was no surprise at all to physicists: it had been predicted by the Higgs mechanism and the standard model of fundamental particle physics almost 50 years earlier.

Naturally, Kuhn associates *that-what* discoveries with scientific revolutions and *what-that* discoveries with normal science, the latter of which seeks to increase the scope and precision of a paradigm by way of puzzle-solving (Kuhn 1962/1996). But as I already mentioned, Kuhn's distinction between the two types of discovery can be used without committing oneself to the entirety of the Kuhnian apparatus, which involves arguably irrational theory change, incommensurability, perceptual changes, world (!) changes, and more. In this paper, I will use this truncated version of Kuhn's account of discovery (see also Schindler 2015).

For readers who disagree that Kuhn's account of discovery can be isolated from other elements of his view are invited to conceive of the account of discovery as self-standing. The purpose of this paper is not exegetical: I would not want to claim that Kuhn himself would subscribe to *every part* of it or that the account captures *all* that Kuhn thought about discovery, but I certainly would still want to describe the account as Kuhnian (see e.g., Okasha (2011) for a similar approach to part of Kuhn's oeuvre). Before articulating hypotheses for the empirical study that I conducted, I will in the next section outline the basic elements of this account of discovery in more detail: the requirement of correctness, the scope of the distinction, and the temporal profile of the two types of discovery.

## 2.1 The Requirement of Correctness

An exegetical detail I *would* indeed stand by is this: Kuhn believed that scientific discoveries of natural kinds require a *correct* conceptualization of the detected phenomenon. For anybody only loosely familiar with Kuhn's philosophy, this statement must appear highly suspect; it requires some explanation.

Let us consider again the discovery of oxygen. Why is Kuhn reluctant to say that Priestley discovered oxygen? After all, it was he who first detected it. Given Kuhn's view of incommensurability, one might have thought that Kuhn would say that Priestley discovered dephlogisticated air in *his* paradigm and that this discovery was somehow replaced in the next paradigm. But Kuhn chose not to say that. On the contrary, Kuhn is pretty clear on Priestley *not* deserving full credit for the discovery because *he held the wrong conception*. Kuhn is even reluctant to call Lavoisier the discoverer of oxygen. Lavoisier successfully isolated oxygen (upon advice by Priestley) and understood that the new gas called for a radically new conception, but he still held a number of false beliefs about oxygen: for him, oxygen gas was a combination of the "principle" of oxygen (apparently modeled on the "principle" of phlogiston) and caloric, the non-existent substance of heat, which would not be abandoned until 1860. Kuhn points out that it would be absurd to say that oxygen was not discovered until then. Still, he does think that the developed concepts of the new phenomenon must be *somewhat* correct. He therefore concludes that oxygen was discovered *sometime* between 1774 (when Priestley first isolated it) and 1777 (when Lavoisier concluded the new gas was oxygen gas), or, as he says, "shortly thereafter" (55).

Hudson (2001) has called out Kuhn on this rather unsatisfactory degree of vagueness concerning the requirement of correctness: as he put it, Kuhn has "left us with the quandary concerning how well one must conceptualise the discovered object" (78). Hudson's own proposal is, roughly speaking, that one only has to have *enough* conceptual resources to successfully identify the new phenomenon in question (Hudson 2001, 77). And for that, the conceptual resources used need not be correct (Hudson 2001, 88). Hudson has no qualms attributing the discovery of oxygen entirely to Priestley. But this does not solve the issue raised by Kuhn: Priestley did not isolate de-phlogisticated air but oxygen. This fact must surely be part any cogent account of the discovery of oxygen.

My own proposal has been that the discovery of natural kind X requires not just an observation of X but also the correct identification of *some* of X's essential properties, namely those properties that suffice to correctly individuate X at a particular time (Schindler 2015). Both theoretical and observational contributions deserve credit. Priestley clearly was the first to observe oxygen. Still, it was Lavoisier, not Priestley, who correctly identified the essential property of oxygen of having a specific mass (in virtue of being a chemical substance). Lavoisier had other beliefs about oxygen that were clearly false, but what he got right about oxygen sufficed to correctly identify it as a new substance at the time. In other words, one need not have a complete description of *all* of the essential properties of X.

Whatever one's own preference for *full* account of scientific, natural kind discovery might be,[2] we can here assume that the Kuhnian account entails a correct, but not necessarily complete, conceptualization of the newly observed phenomenon as an instance of a new natural kind or new property. The hypothesis that suggests itself is therefore:

---

[2] Arabatzis (2006) has argued that in one's analysis of scientific discovery, one need not make any commitment regarding the reality of the discovered phenomena. I am skeptical of this view.

- *Hypothesis 1:* scientific discoveries of a new natural kind (or a new property of a kind) consist of both an <u>observation</u> of an instance of that kind (or property) and an at least partially correct (but not necessarily complete) <u>conceptualization</u> of that instance as belonging to that kind.

A competing hypothesis would be that it is sufficient for a discovery of a natural kind that a new phenomenon be observed, even when the provided conceptualization is not at least partially correct or otherwise faulty (e.g., ad hoc with little independent support).

## 2.2   The Scope of Kuhn's Distinction

Kuhn made clear that he did not believe that *all* scientific discoveries of natural kinds fell into the two types of discovery that we have discussed above (Kuhn 1962, 761, n3). Kuhn mentions the discovery of the positron, as an example, where Anderson first observed the positron in 1932, without much theoretical guidance, and where Blackett and Occhialini at the same time obtained experimental evidence for the positron, using Dirac's equation, which predicted antimatter (see Hanson 1963). It seems that in the hands of Anderson, the discovery was a *that-what* discovery, whereas it was a *what-that* discovery in the hands of Blackett and Occhialini. However, Blackett and Occhialini did not publish their results until after Anderson's paper came out. This is surely relevant to the question of who discovered the positron and what type of discovery this is: only published results are available to the scientific community and thus verifiable. Consequently, it was Anderson, not Blackett and Occhialini, who received a Nobel Prize for the discovery (1936). Of course, the date of publication is contingent and could have been the other way around. Then we may have had to group the discovery as a *what-that* discovery. In either case, a discovery will fall into either of these two classes, unless both kinds of discoveries are published at exactly the same time. I know of no such case.

It is plausible that individual discoveries are connected in longer discovery chains. As we shall see later in this paper, there are examples of *that-what-that* discoveries, where first a new phenomenon is observed, then the phenomenon is conceptualized, and then this new conceptualization leads to predictions of new phenomena which are then observed (see section 4 for examples). Still, we can analyze such a discovery "chains" as a *that-what*-discovery that was followed by a *what-that* discovery. The two types of discovery are thus best conceived of as *minimal units*, which may very well be connected to other discoveries as in the aforementioned way or otherwise.[3] I therefore set out to test the following hypothesis:

- *Hypothesis 2*: scientific discoveries of natural kinds fall into two basic classes: *what-that* discoveries, and *that-what* discoveries.

From the way Kuhn describes the two kinds of discoveries of natural kinds (see above), we can derive:

- *Hypothesis 3*: *what-that* discoveries of natural kinds are <u>expected</u> and *that-what* discoveries are <u>surprising</u> (both relative to the accepted conceptual resources at the time).

Note that these first three hypotheses are clearly *empirically* testable: hypothesis 1 would be wrong if it would suffice for a natural kind discovery to detect a new phenomenon, or, vice versa, if it would suffice for a natural kind discovery to have an incorrect conception of the detected phenomenon; hypothesis 2 would be false if there were only one type of discovery, or if there were several types of discovery, none of which would fall under these two types; and hypothesis 3 would be false if *what-that* discoveries were described of as surprising or *that-what* discoveries as expected.

---

[3] Not all types of discovery chains seem to be possible though. It would seem, for example, that *what-that-what* discoveries are ruled out by the way that Kuhn understood discoveries, as a *what-that* discovery entails that the observed phenomena were predicted; there should be no further need for conceptualization.

There are further hypotheses which one can draw from Kuhn's account. They concern *that-what* discoveries and revolutionary paradigm change, the temporal profile of the two types of discoveries, and the relative importance of observation and theory. I will reserve a section for each in what follows.

## 2.3   Are All *That-What* Discoveries Revolutionary?

Kuhn poses the question of whether all *that-what* discoveries involve paradigm change and answers it hesitantly: "to that question, no general answer can yet be given" (Kuhn 1962/1996, 56). Kuhn then discusses the discovery of X-rays in 1895 by Roentgen and argues that although the discovery did not require a change in paradigm theory, it did bring about a paradigm change with regards to the instrumental and experimental techniques used by scientists when conducting cathode ray experiments (Kuhn 1962/1996, 58-59). Kuhn does not comment further on this question, so it would seem that Kuhn believed that at least most *that-what* discoveries require revolutionary paradigm change *of some sort or another*, be it theoretical, instrumental and experimental, or other. Furthermore, he did not associate *what-that* discoveries with revolutions at all: for him, they were just part of normal science. This gives rise to our fourth hypothesis:

- *Hypothesis 4*: *that-what* discoveries require revolutionary paradigm change, whereas *what-that* discoveries do not.

Of course, Kuhn thought that revolutionary paradigm change entails the incommensurability of paradigms. However, I am not willing to make this commitment; it is also not necessary to recognize paradigm change as revolutionary, i.e., as truly novel, far-reaching, disturbing with regard to previously held views, etc. In this paper, we will use this more modest meaning of "revolutionary". Again, the main purpose of this paper is not exegetical: readers unhappy with this truncation are invited to think of the account used in this paper as self-standing (see also my remarks at the beginning of section 2).

## 2.4   Temporal Profiles and Epistemic Uncertainty

Kuhn attributes different temporal profiles to *that-what* and *what-that* discoveries. Relative to the observation of the new phenomenon, Kuhn thinks of *that-what* discoveries as "necessarily" extended in time, whereas *what-that* discoveries occur in an "instant" (Kuhn 1962, 761-2; 1962/1996, 52-56). That is so, because in *what-that* discoveries scientists are conceptually prepared for what is to come, and when they observe the new phenomenon, the discovery is completed. In contrast, in *that-what* discoveries, when the new phenomenon is first observed, time is required to conceptualize the new phenomenon. It is for this reason that Kuhn also thinks that it is hard to precisely date these kinds of discoveries, and accordingly, why these discoveries are more often associated with priority disputes (Kuhn does not provide any systematic evidence for latter claim) (ibid.).

It is questionable how important this temporal contrast really is to Kuhn's distinction. After all, there is a sense in which *what-that* discoveries are also extended in time; Kuhn's decision to measure the time extension of discoveries relative to the observation of the new phenomenon – rather than relative to the conceptualization – seems arbitrary to some extent. What seems more important to Kuhn's distinction is that at the time of first observation of the new phenomenon, there is *epistemic uncertainty* involved in *that-what* discoveries as to the identity of the discovered phenomenon that is absent in *what-that* discoveries. A relevant hypothesis is therefore this:

- *Hypothesis 5*: after the first observation of a new natural kind, there is greater epistemic uncertainty about its identity in *that-what* discoveries than in *what-that* discoveries.

## 3   Analyzing the Nobel Prizes: Method

In order to test the Kuhnian account of natural kind discovery and the five hypotheses formulated in the previous section, I chose to analyze the Nobel Prizes in Physics in the past 53 years. The Nobel Prize is widely regarded as the highest and most important accolade in the natural sciences. An analysis like the present one should be based on the best and most important bits of science and the Nobel Prizes almost guarantee to highlight such discoveries.

Since Kuhn's preferred field was physics, the Nobel Prizes in Physics were a natural choice for my analysis. I restricted my analysis from the (at the time of the current analysis) most recently awarded Nobel Prize in 2024 to 1972 for a significant practical reason: the text analysis that I carried out was based on documents provided by the Nobel Committee which do not reach back further than 1972 (NobelPrize.org 2025). Also, there are added complications with the earliest Nobel Prizes, such as a heavy bias against any theoretical work until WWI (Friedman 2001). No other systematic biases have been reported after WWI that would be relevant to the current study.[4]

Previous analyses of the Nobel Prizes have focused on the individual scientists receiving the prizes, their age, nationality, and area of expertise (Karazija and Momkauskaitė 2004; Nilesh and Pranav 2018; Bjørk 2019). The current study chose a different unit of analysis, namely the content of the prizes. Each Nobel Prize (NP, for short) is associated with a very short phrase indicating what the prize has been awarded for. The Nobel Committee refers to these phrases as "prize motivation" (they can also be found on the individually designed Nobel diplomas), but, because of their brevity, I will refer to them here as "praise phrases". By the statutes of the Nobel foundation, each prize can be divided between maximally two contributions to science, "each of which is considered to merit a prize". That means that one prize has maximally two praise phrases, each of which I effectively treated as separate prizes (bar phrases summarizing two praise phrases under a common theme, which I ignored).[5] Furthermore, the Nobel Prize in Physics cannot be awarded posthumously, cannot be shared by more than three people, and cannot be awarded to institutions (in contrast to the Nobel Peace Prize).

In the period from 1972 to 2024 there were 53 Nobel Prizes in Physics. Of those, 18 came with two praise phrases each, amounting to 71 praise phrases overall. The current study focused on only praise phrases associated with the discovery of new natural kinds; there were 33 of them. There were three other categories of praise phrases, which I ignored: general theoretical contributions to the field (without specified natural kind), instrumental inventions, and modeling achievements (also without specified natural kind). Table 1 lists these categories and the respective counts of praise phrases:

|  | natural kinds | general theoretical | instrumental | modeling | all |
|---|---|---|---|---|---|
| praise phrase (pp) | 33 | 10 | 27 | 1 | 71 |
| % of all pp | 46% | 14% | 38% | 1% | 100% |

---

[4] The Nobel Prizes seem to be biased against women: even if taking into consideration the relatively low percentage of female scientists in the natural sciences; see Lunnemann et al. (2019). But note that this is not a bias relevant to my analysis: gender is not a variable my analysis is sensitive to. The focus of the current study simply lies elsewhere.
[5] If there are three awardees for a prize, then the prize (and prize money) is either split into three thirds or into one half and two quarters. In the former case, I found, there will be only one praise phrase, and in the latter case, there will be two praise phrases.

*Table 1: Number and percentages (rounded) of praised achievements in the Physics Nobel Prizes, 1972-2024, distinguished by category*

Here is an example of a praise phrase which honors a discovery of a new property of a natural kind, namely the structural duality of collective and individual nucleon motion in atomic nuclei (NP 1975):

- Aage Niels Bohr, Ben Roy Mottelson and Leo James Rainwater … "for the discovery of the connection between collective motion and particle motion in atomic nuclei and the development of the theory of the structure of the atomic nucleus based on this connection".

While in this example, *both* theoreticians (by Rainwater and Bohr) and experimenters (by Mottelson) were honored, in most cases the Nobel Committee bestowed the prize upon *either* experimenters *or* theoreticians: 21 vs. 8 of 33, respectively. For example, the NP 2013 for the discovery of the Higgs boson went only to the theoreticians (although the experimental contributions are also clearly acknowledged in the praise phrase):

François Englert and Peter W. Higgs … "for the theoretical discovery of a mechanism that contributes to our understanding of the origin of mass of subatomic particles, and which recently was confirmed through the discovery of the predicted fundamental particle, by the ATLAS and CMS experiments at CERN's Large Hadron Collider".

Only in four cases did the Nobel Committee award the Prize to both experimentalists and theoreticians for the same natural kind discovery (as in the case of the NP 1975, mentioned above). Whether or not the Committee conferred the prize to experimenters, theoreticians, or both, was *not* part of my analysis of natural kind discoveries (however I will come back to the issue of credit distribution in the final section of this paper).

Let us now consider examples of instrumental (26), modelling (1), or theoretical achievements with no particular natural kind discovery, but rather more general theoretical contributions to the field (10):

- *Instrumental*: e.g., Bertram N. Brockhouse "for the development of neutron spectroscopy"; Clifford G. Shull "for the development of the neutron diffraction technique" (NP 1994);
- *Modelling*: e.g., Syukuro Manabe and Klaus Hasselmann "for the physical modelling of Earth's climate, quantifying variability and reliably predicting global warming" (NP 2021).
- *Theoretical; no particular observation*: e.g., James Peebles "for theoretical discoveries in physical cosmology" (NP 2019);

For all discoveries whose Nobel Prize praise phrase mentioned the discovery of a new natural kind or property, I recorded the years in which the observations pertaining to a discovery were made (the "that"), and when the theoretical understanding of it was developed (the "what"), if at all. The Kuhnian account of discovery would lead us to expect a list of *what-that* and *that-what* discoveries to result from this (see hypothesis 1 and 2). In order to see how the discoveries were described ("surprising", "expected", or "revolutionary"), and whether hypothesis 3-4 were correct, I then conducted a text analysis of the press releases and explanatory documents which the Nobel foundation provides on its website back until the NP 1972 (NobelPrize.org 2025). For this analysis, the text associated with each discovery was hand-coded. The results were cross-validated by an independent coder. The test of hypothesis 5 was a bit more involved and will be explained later in the results section.

It turned out that there were several complications in categorizing discoveries by simply recording the observation of a new phenomenon and its correct conceptualization. This became clear from a closer study of the texts accompanying the discoveries and from the way the discoveries were described. For the remainder of this section, I will explain these complications. Readers not interested in these complications may now skip to the results discussed in section 4.

Most importantly, perhaps, there were discoveries where a theory was developed before the observations, but where the theory *so happened* not to guide the experiments. This, for example, was the case for the first compelling evidence for the existence of quarks. James Bjorken predicted the scaling behavior of subatomic (hadronic) particles in 1967, but the MIT-SLAC deep inelastic scattering experiments of 1968 unexpectedly discovered scaling and *then* sought Bjorken's help for interpreting the results (Riordan 1992). For our purposes, the discovery thus counts as a *that-what* discovery. The Nobel prize of 1990 went to the experimentalists (Bjorken never received a Nobel prize). Other examples in which the relevant theory was available to the experimenters in principle, but not used, concern Arno A. Penzias and Robert Woodrow's discovery of cosmic microwave background in 1965 (NP 1978),[6] and Michel Mayor and Didier Queloz's 1995 discovery of the first exoplanet, called "51 Pegasi b", for which they received the NP 2019.

There were discoveries which proved to be deceptive. Again, Mayor and Queloz's discovery of "51 Peg" is such a case. Their discovery may at first seem to be a straightforward observation of planetary motion in just another solar systems. So, one may think that this discovery should be grouped as a *what-that* discovery. However, it turns out that the discovery was in fact unexpected and required some real theoretical work: 51 Peg has the size of Jupiter but is surprisingly close to its sun. Many astronomers thus at first suspected that the observations were caused by other effects such as stellar pulsation and star spots. It had to be argued on theoretical grounds that 51 Peg had actually formed at much larger distances away from its sun (around 5 AU instead of 0.5 AU) and then migrated toward the sun, before it could be accepted that the first exo-planet had been discovered (NobelPrize.org 2025).[7] It is clear from the documents provided by the Nobel Foundation that Mayor and Queloz initially experienced some significant resistance in getting their discovery recognized as an exoplanet (see below section 4.3). I therefore categorized this discovery as a *that-what* discovery.

There were also discoveries that seemed to combine two types of discovery in one award, and discoveries that spread over several awards. Let us begin with the first kind of case, which can be illustrated with the NP 1975, which, as we already mentioned, was awarded to Bohr, Mottelson, and Rainwater for their work on models of the atomic nucleus. Rainwater's work was prompted by observations of aberrations from spherical symmetry in the charge distribution of certain atomic nuclei – aberrations known as electric quadrupole deformations – which were at odds with the existing liquid drop models. Rainwater then proposed that this observed asymmetry arose from an asymmetry in the distribution of nucleons within the nucleus. Rainwater's contribution may thus be thought of as a contribution to a *that-what* discovery. Rainwater's ideas were then further developed by Bohr into a full-fledged theory of collective motion in deformed nuclei, which was then confirmed experimentally by Mottelson. This part of the discovery should clearly count as a *what-that* discovery. Because the Nobel Committee opted not to give the award to the observations prompting Rainwater's theoretical work, and because the prize went to Bohr for his theory and Mottelson's confirmation of the theory instead, I decided to categorize the discovery as a *what-that* discovery.

---

[6] Penzias and Woodrow's paper appeared alongside a paper by Dicke, Peebles, and colleagues which showed that the detected microwave background could be accounted for by the residual thermal radiation from the early universe. The big bang theory was developed already in the 1940s but was not taken too seriously until Penzias ad Woodrow's results.
[7] I suggest thinking of the discovery of exoplanets as a discovery of a new property of the known natural kind of planets. For a detailed discussion of planets as natural kinds see Magnus (2012).

There are discoveries which have spread over several prizes. For example, CP symmetry violation was first observed in kaon decay in 1964 by James W. Cronin and Val L. Fitch and explained in 1973 by Makoto Kobayashi and Toshihide Maskawa, who integrated symmetry breaking into the standard model by introducing a third quark generation (namely, bottom and top). Their model was supported by the discovery of this third generation of quarks (1977 and 1995) and fully confirmed by experiments measuring CP-violation in B-meson decays in 2001 as predicted by the KM model (BaBar at Stanford and Belle at Tsukuba, Japan). Clearly, the episode could be described as a discovery "chain" (namely a *that-what-that* discovery), but this chain is still constituted by the two basic types of discovery: a surprising *that-what* discovery (for which the experimentalists Cronin and Fitch received the NP 1980) and a *what-that* discovery resulting from a prediction of the KM model (for which the theorists Kobayashi and Maskawa received the NP 2008).

There was another example that followed a similar pattern, namely the discovery of solar neutrinos, which resulted in two Nobel Prizes (2002 and 2015). Neutrinos were already postulated in the 1930s, but it was not until the 1950s and 1960s that models of the expected flux of solar neutrinos were developed by John Bahcell. The famous Homestake experiments led by Raymond Davis – and later also the Kamiokande experiments led by Masatoshi Koshiba – then detected a significant discrepancy between the observations and the predictions by these models (by not less than 1/3). This discrepancy was later explained in terms of neutrino oscillations.[8] Intriguingly, Davis and Koshiba did not receive the NP 2002 before this explanation had been confirmed in experiments by Takaaki Kajita and Arthur B. McDonald in 2001, for which they in turn received the NP 2015. The episode can again be described as a discovery *chain* consisting of a *that-what* discovery and a *what-that* discovery.

# 4  Results and Discussion

In this section I will report the results I obtained from my analysis of the Nobel Prizes in Physics from 1972 to 2024 with a view of testing Kuhn's account of scientific discovery. I will first report the results relevant to hypothesis 1 and 2 (section 4.1), hypothesis 3 and 4 (section 4.2), and hypothesis 5 (section 4.3).

## 4.1  Hypothesis 1 and 2: Correct Concepts and the Two Types of Discovery

The results confirm hypothesis 1: for the period I looked at, there was only a single discovery which received the Nobel Prize *before* it was *correctly* understood *what* the phenomenon was that was discovered. There were no prizes that were made before an at least partially correct (and not necessarily complete) conception of the new phenomenon had been provided. Remarkably, none of the developed conceptions later turned out to be incorrect or incomplete, except one.[9]

Hypothesis 2 was also fully confirmed: all of the natural kind discoveries that I identified fall into the classes of *what-that* discoveries and *that-what* discoveries. More specifically, the study found 19 *that-what* discoveries and 14 *what-that* discoveries. As *what-that* discoveries I counted (parts of) the Nobel Prizes of the following years: 2022, 2020, 2017, 2015, 2013, 2008, 2006, 2001, 1995, 1993, 1984-5, 1979, 1975. As

---

[8] According to this explanation, electron-neutrinos coming from the sun turn into tau- or muon-neutrinos; only electron neutrinos had been detected by the earlier experiments. The undetected muon-neutrinos accounted for the discrepancy between theory and experiment.

[9] The exception concerns the discovery of high temperature superconductors in 1986 (NP 1987) that went to the experimentalists Bednorz and Mueller. Philip W. Anderson proposed an explanation of these results in terms of resonating valance bond theory, but the explanation is now widely regarded as not fully adequate. It is possible that the Nobel Prize Committee was swayed by Anderson's explanation *at the time*. Also, and perhaps more importantly, since the discovery is an extension of the well-known and explained phenomenon of superconductivity, the Nobel Committee may have applied lower demands on the conceptualization of the new phenomenon than in other discoveries.

*that-what* I counted the years in: 2019, 2011, 2005, 2003(2x), 2002, 1998, 1995-96, 1990, 1988, 1987, 1982, 1980, 1978, 1976, 1972-74. See the appendix at https://osf.io/wyf2q/ for a complete list of discoveries.

## 4.2 Hypothesis 3 and 4: Discovery Predicates

The way that the discoveries were described on www.nobelprize.org – namely as "expected", "surprising", or revolutionary – matched their categorization as *that-what* or *what-that* discovery almost perfectly (see table 5).[10] The cross-validation by an independent coder obtained an excellent match (Cohen's Kappa = 0.862).[11] I consider hypothesis 3 to be confirmed.

|  | surprising | expected | revolutionary |
|---|---|---|---|
| *that-what* (19) | 100% (19) | 0 | 74% (14) |
| *what-that* (14) | 7% (1) | 93% (13) | 50% (7) |

*Table 2: Nobel Prize descriptions. The table shows descriptions associated with the two types of discovery. Percentages are relative to the total number of discoveries of the respective kind.*

The analysis also revealed that *that-what* discoveries were more frequently described as revolutionary than *what-that* discoveries (74% vs. 50%). The intercoder cross-validity was again excellent (Cohen's Kappa = 0.832). This result is ambiguous with regards to hypothesis 4: although the difference seems to accord with Kuhn's view that *that-what* discoveries are revolutionary, there is still a quarter of *that-what* discoveries that were *not* described as revolutionary. While this may be blamed on poor description in the analyzed documents, it's perhaps more significant that still half of all *what-that* discoveries are described as revolutionary too. This is clearly not how Kuhn thought about these types of discoveries.

## 4.3 Hypothesis 5: Epistemic Uncertainty

As mentioned in section 2, on the Kuhnian view of discovery, there is more epistemic uncertainty about the identity of the discovered phenomenon in *that-what* discoveries than there is in *what-that* discoveries (this is hypothesis 5). One way in which this epistemic uncertainty could express itself is in the temporal distance between the observation of the new phenomenon and its first correct conceptualization: one might expect that distance to be larger in *that-what* discoveries than in *what-that* discoveries. Table 3 shows that this is not at all the case: whereas for *that-what* discoveries the median temporal distance is 1 year, it is 38 years for *what-that* discoveries (table 3).[12]

|  | Time lag between *that* and *what*: **median** (and mean) |
|---|---|
| *that-what* | **1** (13.1) |
| *what-that* | **38** (26.9) |

*Table 3: Median (and mean) temporal distance in years between observation and conceptualization. The table indicates the median (and mean) time passed (in years) between the observation and conceptualization (in that-what discoveries) and vice versa (in what-that discoveries).*

---

[10] The only exception is the NP 1985 for the discovery of the quantized Hall effect. Although it had been expected before the discovery that conductivity of two-dimensional materials changes in a step-wise function, the immense precision with which quantum rules applied was very much surprising.

[11] Intercoder reliability is considered excellent when Cohen's Kappa ≥ 0.8, see e.g., O'Connor and Joffe (2020).

[12] Occasionally, no precise date of observation or conceptualization could be determined. The Homestake Experiment detecting solar neutrinos that ran from the 1960s to the 1990s and the explanation of cosmic microwave background by the big bang theory (that was developed over decades) are cases in point. In these cases, averages of the respective time ranges were used.

At first glance, one may think that this result contradicts Kuhn's view that *that-what* discoveries "necessarily" take time whereas *what-that* discoveries happen in "an instant" (see section 2.4): here it seems to be the reverse. Yet, we must remind ourselves that Kuhn makes this statement relative to the first observation of the new phenomenon. So, the time that passes between the prediction and the first observation in *what-that* discoveries is irrelevant to this claim. I therefore conducted another test, which I will get to after commenting on this result in a bit more detail.

An obvious explanation of the difference recorded in table 3 is that the required observations can be hard to come by: it can be technologically challenging to build and conduct the experiments necessary for the confirmation of theoretical predictions (in *what-that* discoveries). The discovery of the Higgs boson, which required millions of euros and many years to build the Large Hadron Collider, is a case in point. Not all experiments are as difficult, but there could be a pre-selection of especially arduous experiments for the Nobel Prizes, because it might be precisely those kinds of *what-that* discoveries that the Nobel committee finds prize-worthy.

By comparison, developing the right concepts can be cheap, as is evidenced by the small temporal difference between the observation and the right concept. There were in fact six discoveries, in which the right concept was developed in the same year as the observation. Then again, there are also cases in which it took a very long time to come up with the right ideas: as already mentioned, it took 46 years for superconductivity to receive the correct theoretical explanation (NP 1972). So, while the development of new concepts is not guaranteed to be relatively fast, compared to experimental confirmations it is on average much faster.

The other way that I went about testing hypothesis 5 was by measuring the temporal distance between the time when a discovery was completed and *the time when the Nobel Prize was awarded*. The underlying idea was that more epistemic uncertainty would result in the Nobel Committee taking more time to award a Prize. Since Kuhn attributes greater epistemic uncertainty about the new phenomenon to *that-what* discoveries, it should be these discoveries for which the Nobel Committee should take relatively more time to make an award. This prediction was borne out by the results (see table 4): it took the Nobel Committee *more than twice as long* to award the prize to *that-what* discoveries as it did to award it to *what-that* discoveries. I take this to confirm hypothesis 5.

|  | *that* until prize | *what* until prize |
|---|---|---|
| *that-what* | 20 (30.7) | **15.5** (17.6) |
| *what-that* | **7** (8.2) | 52 (47.6) |

Table 4: The two discoveries and the median (and average) temporal distance from the completion of the discovery to the award of the prize (in years)

The reason that suggests itself for this difference is that it takes the community more time to assess whether the conceptualization is appropriate in *that-what* discoveries, whereas in *what-that* discoveries, the community has had ample time to do so before the observation was finally made. This explanation is supported also by the descriptions that can be found in the documents provided by the Nobel Foundation (NobelPrize.org 2025). Consider for example an excerpt from the Nobel lecture by Didier Queloz for his discovery of the first exoplanet, for which he (and Michael Mayor) received the Nobel Prize in 2019:

> In the following years we would be confronted with a wave of skepticism. It would take years for the community at large to accept the reality of 51Peg hot Jupiter and to modify the paradigm about the universality of solar system planetary architecture […] The main issue was that it didn't fit in the planetary formation paradigm without seriously tweaking this paradigm. Changing a well-established theory is rarely the first idea a physicist is considering out of an unusual experimental result. And yet the foundation of planet formation theory needed to be revised. (Queloz 2019, 121-122)

Clearly, there was epistemic uncertainty about the identity of the newly observed phenomenon. Comments like these are not at all untypical for *that-what* discoveries. Consider, for example, also Martin Perl's Nobel Lecture for the discovery of tau muons, for which he received the NP 1995. In a section entitled "Is it a lepton?", Perl writes that "our first publication was followed by several years of confusion and uncertainty about the validity of our data and its interpretation", among other things, because theory did not require a third charged lepton and there were also otherwise no reason to expect a third lepton (Perl 1995, 186-187). Establishing the identity of the new phenomenon also requires ruling out all sorts of competing interpretations of experimental results (Perl 1995, 187-188).

## 5  Conclusion and Normative Implications

The last 53 years of Nobel Prizes suggest that Kuhn's distinction between two types of natural kind discovery is indeed an insightful one to make. According to Kuhn, any scientific discovery of natural kind X requires not only an observation of X but also a correct conceptualization of X. This is clearly evidenced by the Nobel Committee refraining from awarding prizes for the discovery of phenomena for which no adequate understanding had yet been developed; except one, none of them turned out to be incorrect.

The current analysis also found that all natural kind discoveries awarded with a Nobel Prize fell into either the category of a *what-that* discovery, or a *that-what* discovery. Most significantly, perhaps, in the material provided by the Nobel Foundation *that-what* discoveries are consistently described as "surprising" and *what-that* discoveries as "expected". The current study also found evidence for the view that *that-what* discoveries are associated with much greater epistemic uncertainty. While *that-what* discoveries were more frequently described as "revolutionary" than *what-that* discoveries, still half of the latter were described as revolutionary too. This result is clearly contrary to what Kuhn thought, but, in my view, does not fundamentally undermine the usefulness of the distinction, especially if one takes some distance to several elements in the Kuhnian package of ideas (as this paper did).

The Kuhnian distinction between the two types of natural kind discovery allows us to bring order to an otherwise unstructured set of discoveries. But it does more than that. It also has *normative implications*. In particular, if the Kuhnian picture of scientific discovery is true and a discovery of a new natural kind or its properties requires *both* an observation and a theoretical understanding of it, then recognizing only one of these elements would clearly not do justice to a discovery; it also would not do justice to the discoverers. As we saw, the way that the Nobel Prizes have been awarded suggests that the Nobel Committee at least implicitly embraces the Kuhnian view of discovery: again, the prizes are not awarded for observations of new phenomena whose identity is not yet understood; all natural kind discoveries fall into Kuhn's two classes with their distinct characteristics; and *that-what* discoveries must await their prizes longer than *what-that* discoveries, indicating greater epistemic uncertainty also on part of the Committee. And yet, despite the Committee's espousal of the Kuhnian view, its decisions concerning which *contributions* to highlight in their awards, and whether to award the prize to theoreticians or experimentalists, are rather inconsistent and exhibit a strong bias for experimental contributions. Both of these aspects are problematic. Let us consider these two issues briefly in turn.

Sometimes the Committee rewards only theoretical contributions, sometimes only experimental contributions, and only occasionally both, without apparent reason. For example, the NP 1984 for the discovery of the W and Z particles went only to the experimentalists (Rubbia and van der Meer), and the NP 2013 for the discovery of the Higgs boson went only to the theoreticians (Higgs and Englert), even though they are both *what-that* discoveries in high energy physics.[13] It would have been open to the Committee to reward the other element of the discovery (theoretical or experimental) as well by recognizing a third scientist in each of these cases, and maybe this is what the Committee should have done. Another salient case is the discovery of nuclear fission, for which only Otto Hahn received the Nobel Prize, and not Lise Meitner, whose theoretical understanding of the detected physical processes was absolutely crucial to understanding them as fission in the first place. She too – it is widely agreed – should have received a Nobel Prize.[14]

These are not exceptional cases. As table 5 shows, only a small fraction of all *that-what* and *what-that* discoveries were discoveries in which *both* the experimenters and theoreticians were awarded with a Nobel Prize (11% and 14%, respectively). Table 5 also shows that the Committee clearly sports a preference for experimental contributions for both types of discovery.[15]

|  | *that-what* | | *what-that* | |
|---|---|---|---|---|
| both, experimenters and theoreticians: | 2 | 11% | 2 | 14% |
| experimenters only: | 12 | 63% | 9 | 64% |
| theoreticians only: | 5 | 26% | 3 | 21% |

*Table 5: Awarded contributions with respect to the two types of natural kind discoveries. The table indicates the number and percentages of the types of contributions to the Nobel Prizes that were recognized by the Nobel Committee. Percentages are relative to the overall number of discovery type (that-what: 19, or what-that: 14). Percentages are rounded.*

Even when taking into consideration the fact that the Nobel Committee does not award Nobel Prizes posthumously, and assuming that the Committee might have awarded each of the deceased discoverers, this would bring us only up to 32% of *that-what* discoveries and 36% of *what-that* discoveries, for which both contributions might have been awarded.[16] The practice would thus still be rather far off the ideal.

One might object: why think that the Nobel Committee's credit distribution practice is not ideal, instead of concluding that it the Kuhnian account that is, after all, wrong?[17] Note though that facts about *award*

---

[13] Note that the vast number scientists involved in the collaborations carrying out high energy physics experiments does not necessarily constitute an obstacle to awarding the Nobel Prize to such collaborations. As the NP 1984 and other NPs illustrate (e.g., the NP 2017 for the discovery of gravitational waves), the Committee has sometimes chosen to give the award to the leader of such collaborations.

[14] This particular case has also been discussed as a case of sexism in science, which it probably was too (see Sime (1996)). The account proposed here renders the injustice even more infuriating, as it recognizes Meitner's theoretical contribution as necessary.

[15] Karazija and Momkauskaitė (2004) report that in the first ten decades of the Nobel Prize in Physics (1901 to 2000), on average twice as many prizes have gone to experimenters than to theoreticians. This rate is matched perfectly by my numbers: 7 vs. 14 (that-what) and 5 vs. 11 (what-that).

[16] In the range of prizes studied for this paper, there were several cases in which either the theorists or the experimenters had deceased before the Nobel Committee made their awards: there were three *what-that* discoveries in which the award went to experimenters when the theorists had already deceased (NP 2022, NP 2001, and NP 2017) and there are four *that-what* discoveries in which the theorists got the award when the experimenters had already deceased (NP 2005, NP 2003, NP 1982, and NP 1972).

[17] I owe this objection to an anonymous reviewer for this journal.

*emphasis* – namely whether the prize should go to theoretical or experimental contributions – are not facts that could possibly undermine the facts obtained in support of the Kuhnian account concerning natural kind discovery. Those latter facts, recall, are facts about whether a correct conception of the newly observed phenomena had been found at the time of the award (independently of whether that contribution received a prize), whether the discoveries fall into the two classes identified by Kuhn, and whether they have the characteristics assigned to them by Kuhn. In other words, facts about award emphasis and facts about natural kind discovery are evidentially orthogonal.

While the current study presented evidence that both experimental and theoretical contributions are required for a natural kind discovery (in accordance with the Kuhnian view), the current study in no way excludes the possibility of self-standing theoretical, experimental, or other kinds of discoveries. As far as I'm concerned, one may even *speak* of the contributions to a natural kind discovery of X as theoretical and experimental *discovery* of X, so long as one appreciates that *both* are required for natural kind discovery.

The current study has limitations. First, one may argue that the discoveries highlighted by the Nobel Prize must satisfy the highest standards on discovery, and that these standards for discovery may be much weaker for science that is getting less limelight. Whether or not this is the case, cannot be determined here. The second limitation is fairly obvious: the current study focused on physics, and only on a subset of all Nobel Prizes awarded in this area. It would be interesting to see to what extent Kuhn's distinction also applies in the other scientific disciplines in which the Nobel Prize is awarded for natural kinds.

I should emphasize, at last, that the *primary* goal of this paper is not to criticize any specific award practices. Simply in virtue of the very small number of awards and its limitation to three awardees, it is quite obvious that the Nobel Prizes will never be able to do full justice to the complexity of discovery. Rather, the explication of the concept of the discovery of phenomena in science offered here – beyond providing insight into the nature of discovery of natural kinds – is meant to motivate broader reflection on the question of how *in general* credit ought to be distributed in science, which I think is thoroughly needed.

**Acknowledgements:**

I wish to thank two anonymous reviewers of this journal for their very helpful feedback. Thanks also to the audiences at the Bergen Workshop for the Philosophy of Science (2024), the Hong Kong University of Science and Technology, The Epistemology of Experimental Discovery Workshop at the University of Bristol (2025), and the Centre for Science Studies at Aarhus University.

**Funding Statement:** None to declare.

**Declarations**: None to declare.